\documentclass[amstex,prd,aps,graphicx,epsf,tighten]{revtex4}

\newcommand{\bra}{\left\langle}
\newcommand{\ket}{\right\rangle}
\newcommand{\D}{{\mathcal D}}
\newcommand{\Tdot}{\frac{\partial}{\partial t}}

\newcommand{\bo}{{\mathcal{O}}}

\newcommand{\ltb}{LTB\ }
\newcommand{\vpc}{VPC}
\newcommand{\frw}{FLRW\ }

\begin{document}
\title{Averaging in Spherically Symmetric Cosmology}
\author{A.A. Coley\dag ~and N. Pelavas\dag}
\address{\dag\ Department of Mathematics and Statistics,\\
Dalhousie University, Halifax, Nova Scotia, Canada}

\begin{abstract}

The averaging problem in cosmology is of fundamental importance. When applied to study cosmological evolution, the
theory of macroscopic gravity (MG) can be regarded as a long-distance modification of general relativity. In the
MG approach to the averaging problem in cosmology, the Einstein field equations on cosmological scales are
modified by appropriate gravitational correlation terms. We study the averaging problem within the class of
spherically symmetric cosmological models. That is, we shall take the microscopic equations and effect the
averaging procedure to determine the precise form of the correlation tensor in this case. In particular,  by
working in volume preserving coordinates, we calculate the form of the correlation tensor under some reasonable
assumptions on the form for the inhomogeneous gravitational field and matter distribution. We find that the
correlation tensor in a Friedmann-Lema\^{\i}tre-Robertson-Walker (FLRW) background must be of the form of a
spatial curvature. Inhomogeneities and spatial averaging, through this spatial curvature correction term, can have
a very significant dynamical effect on the dynamics of the Universe and cosmological observations; in particular,
we discuss whether spatial averaging might lead to a more conservative explanation of the observed acceleration of
the Universe (without the introduction of exotic dark matter fields). We also find that the correlation tensor for
a non-FLRW background can be interpreted as the sum of a spatial curvature and an anisotropic fluid. This may lead
to interesting effects of averaging on astrophysical scales. We also discuss the results of averaging an
inhomogeneous Lema\^{\i}tre-Tolman-Bondi solution as well as calculations of linear perturbations (that is, the
backreaction) in an FLRW background, which support the main conclusions of the analysis.
\end{abstract}

 \noindent [PACS: 98.80.Jk,04.50.+h] \vskip1pc

\maketitle

\section{Introduction}

The Universe is not isotropic or spatially homogeneous on local
scales. The correct governing equations on cosmological scales are
obtained by averaging the Einstein equations of general relativity
(GR). An averaging of inhomogeneous spacetimes can lead to
dynamical behavior different from the spatially homogeneous and
isotropic Friedmann-Lema\^{i}tre-Robertson-Walker (FLRW) model
\cite{Ellis:1984}; in particular, the expansion rate may be
significantly affected \cite{Bild-Futa:1991}. Consequently, a
solution of the averaging problem is of considerable importance
for the correct interpretation of cosmological data.
Unfortunately, this is a very difficult problem.

There are a number of theoretical approaches to the averaging problem
\cite{Bild-Futa:1991,Zala,MZ,aver,buch,BucCar,Futa}. In the approach of Buchert \cite{buch},  a 3+1 cosmological
spacetime splitting that depends on the spacetime foliation is utilized and, in addition, only scalar quantities
are averaged (and hence, in general, the equations are not closed and consequently additional assumptions are
necessary). The perturbative approach \cite{Bild-Futa:1991,Futa} involves averaging the perturbed Einstein
equations; however, a perturbation analysis cannot provide detailed information about the averaged geometry.

In all of these approaches an averaging of the Einstein equations is performed to obtain the averaged field
equations. The macroscopic gravity (MG) approach to the averaging problem in cosmology is an attempt to
\cite{Zala} give an exact (and tensorial) prescription for the correlation functions which inevitably emerges in
an averaging of the field equations (without which the averaging simply amounts to definitions of the new averaged
terms). In MG the Einstein equations on cosmological scales with a continuous distribution of cosmological matter
are modified by appropriate gravitational correlation (correction) terms. The MG approach provides a covariant
method of averaging tensors, consequently it does not rely on assumptions regarding the nature of perturbations
(e.g., in principle there are no approximations and no higher order terms are dropped). We shall adopt the  MG
averaging approach in this paper.

The spacetime averaging procedure adopted in MG is based on the concept of Lie-dragging of averaging
regions\footnote{In this paper, averaging is performed in one coordinate patch of a volume preserving coordinate
system and so the issue of Lie-dragging of averaging regions does not apply.}, which makes it valid for any
differentiable manifold with a volume $n$-form, and it has been proven to exist on an arbitrary Riemannian
spacetime with well-defined local averaged properties \cite{Zala}. Averaging of the structure equations for the
geometry of GR leads to the structure equations for the averaged (macroscopic) geometry and the definitions and
the properties of the correlation tensor. The averaged Einstein equations for the macroscopic metric tensor
together with a set of algebraic and differential equations for the correlation tensors become a coupled system of
the macroscopic field equations for the unknown macroscopic metric, correlation tensor, and other objects of the
theory. The averaged Einstein equations can always be written in the form of the Einstein equations for the
macroscopic metric tensor when the correlation terms are moved to the right-hand side of the averaged Einstein
equations, and consequently can be regarded as a geometric modification to the averaged (macroscopic) matter
energy-momentum tensor \cite{Zala}.

MG is a non-perturbative geometric field theory with a built-in scale related to the spatial scale over which
averages are taken (we recall that, in principle, averaging is performed over a 4-volume region). The microscopic
field to be averaged is supposed to have two essentially different variation scales, $\lambda $ and $L_H$,
satisfying $\lambda <<L_H$, where $L_H$ is the horizon size related to the inverse Hubble scale. An averaging
region must be taken of an intermediate size $L$ such as $ \lambda <<L<<L_H $, so that the averaging effectively
smooths out all the variations of the microscopic field of the scale $\lambda $. In cosmological applications
$\lambda$ is taken to be the scale on which astrophysical objects such as galaxies or clusters of galaxies have
structure, and the size of the averaging space regions has been tacitly assumed to be $\simeq $\ $100$\ Mpc, or a
fraction of the order of the inverse Hubble scale, and thus any terms (e.g., a cosmological constant or a
curvature term) appearing in the correlation tensor might be expected to be related to the inverse Hubble scale.
In principle the scale, given by the size of the spacetime averaging region, is a free parameter of the theory.

A procedure for solving the MG equations with one connection correlation tensor was discussed in \cite{CPZ}. The
macroscopic field equations were written in the form of the Einstein equations of GR, with a `modified'
stress-energy tensor consisting of the averaged microscopic stress-energy tensor $\langle {\bf t}^{
(micro)}\rangle$ and an additional effective stress-energy tensor ${\bf C}$ arising from the correlation tensor
${\bf Z}$ \cite{Zala}. In \cite{CPZ} it was found that the averaged Einstein equations for a spatially
homogeneous, isotropic macroscopic spacetime geometry has the form of the Einstein equations of GR for a spatially
homogeneous, isotropic spacetime geometry with an additional spatial curvature term (i.e., the correlation tensor
${\bf C}$ is of the form of a spatial curvature term).

Therefore, assuming spatial homogeneity and isotropy on largest
scales, then the inhomogeneities affect the dynamics through
correction terms (the correlation tensor) of the form of a
curvature term \cite{CPZ}, which will dominate at late times and
on largest scales. Thus even for FLRW backgrounds it is important
to understand how these correction terms affect cosmological
observations. For example, a spatially averaged metric is not a
local physical observable: the averaged value of the expansion
will not be the same as the expansion rate of the averaged
geometry, because of the non-linear nature of the expansion.

The spacetime averaging in MG utilizes bilocal averaging operators. The MG averaging scheme is especially simple
in a proper coordinate system \cite {MZ}, in which the bilocal operators takes on the simplest possible forms. In
particular, any proper coordinate system is necessarily a volume-preserving (system of) coordinates (VPC), and in
a pseudo-Riemannian spacetime the spacetime averages defined in proper coordinates are Lorentz tensors exactly
like the averages in Minkowski spacetime; that is, VPC on an arbitrary differentiable metric manifold is a natural
counterpart of the Cartesian coordinate system on a Minkowski manifold. A brief review of the spacetime averaging
scheme adopted in macroscopic gravity and the role of proper systems of coordinates is presented in Appendix A.

Spherically symmetric cosmological models are of special
cosmological importance, partially motivated by the observed
isotropy of the CMB. Therefore, it is important to study the
averaging problem comprehensively within the class of spherically
symmetric cosmological models (i.e., to determine the form of the
MG equations in the case of spherical symmetry). We shall take the
microscopic equations and effect the averaging procedure to
determine the precise form of the correlation tensor in this case.

In the next section we shall calculate the form of the MG equations in the case of spherical symmetry. The first step
is to choose an appropriate spherically symmetric VPC system. It is also instructive to investigate the FLRW metric in
VPC. We then make some reasonable assumptions on the form of the inhomogeneous gravitational field and matter
distribution, and in section III we calculate the resulting form of the correlation tensor in both a FLRW and non-FLRW
background.

In section IV we average an inhomogeneous Lema\^{\i}tre-Tolman-Bondi (LTB) solution. The first step in this
calculation is to rewrite  the LTB dust solution in VPC (which is done in Appendix B). This solution is presented
as a perturbation about a flat \frw model. In section V we then assume a spatially homogeneous and isotropic
background and discuss the effect of perturbations (that is, the backreaction) on this FLRW background.

In section VI we discuss the results obtained in light of recent observations, with particular emphasis on the effect
of inhomogeneities on the local expansion rate. We discuss whether inhomogeneities and spatial averaging might lead to
a more conservative explanation of the observed acceleration of the Universe (without the introduction of exotic dark
matter fields). The conclusions are given in the final section. In Appendix C we briefly discuss the relationship
between our work and the work of Buchert \cite{buch} in the case of spherical symmetry.

\section{Spherical Symmetry}
We shall calculate the form of the MG equations in the case of
spherical symmetry.  That is, we shall take the microscopic
equations and effect the averaging procedure to determine the
precise form of the correlation tensor, $C^{a}_{\ b}$, in this
case.

We begin by choosing an appropriate coordinate system.  Starting from the general form of the spherically symmetric
metric, we first choose a new angular coordinate, $u=cos(\theta)$, to eliminate any angular dependence in $\sqrt{-g}$,
where $g=det(g_{ab})$.  Next, we use the remaining coordinate freedom to set $\sqrt{-g}=1$; this is done by choosing an
appropriate form for the `radial' metric function that multiplies the spherical line element $ds^{2}(u,\phi)$. The line
element is thus \footnote{In general, these are not comoving coordinates.}

\begin{equation}
ds^2=-Bdt^2+Adr^2+\frac{du^2}{\sqrt{AB}(1-u^2)}+\frac{1-u^2}{\sqrt{AB}}d\phi^2,
\label{vpcss}
\end{equation}

\noindent where the functions $A$ and $B$ depend on $t$ and $r$.
These are volume preserving coordinates (VPC) for the spherically
symmetric metric \cite{MZ}.  It is the adoption of VPC that
enables us to calculate the averaged quantities in a relatively
straightforward manner.

We next calculate \cite{grtensor} the form of the Einstein tensor $G^{a}_{\ b}$
(note the position of indices)

\begin{eqnarray}
G^{t}_{\ t} & = & \frac{5}{8}\,{\frac {B_{{r}}A_{
{r}}}{{A}^{2}B}}-\frac{1}{2}\,{\frac
{B_{{rr}}}{AB}}+\frac{3}{16}\,{\frac {{A_{{t}}}^
{2}}{{A}^{2}B}}+\frac{1}{8}\,{\frac
{A_{{t}}B_{{t}}}{A{B}^{2}}}+{\frac {11}{16 }}\,{\frac
{{B_{{r}}}^{2}}{A{B}^{2}}}-\frac{1}{16}\,{\frac
{{B_{{t}}}^{2}}{{B} ^{3}}}-\frac{1}{2}\,{\frac
{A_{{rr}}}{{A}^{2}}}-\sqrt {AB}+{\frac {15}{16}}\,
{\frac{{A_{{r}}}^{2}}{{A}^{3}}} \label{Gtt} \\
G^{r}_{\ t} & = & -\frac{5}{8}\,{\frac {B_{{r}}A_
{{t}}}{{A}^{2}B}}-{\frac {7}{8}}\,{\frac
{B_{{r}}B_{{t}}}{A{B}^{2}}} -{ \frac {7}{8}}\,{\frac
{A_{{r}}A_{{t}}}{{A}^{3}}}-\frac{1}{8}\,{\frac {A_{{r}}B
_{{t}}}{{A}^{2}B}}+\frac{1}{2}\,{\frac{A_{{rt}}}{{A}^{2}}}+\frac{1}{2}\,{\frac
{B_{{ rt}}}{AB}} \\
G^{r}_{\ r} & = & -\frac{5}{8}\,{\frac {A_{{t}}B_
{{t}}}{A{B}^{2}}}+\frac{1}{2}\,{\frac {A_{{tt}}}{AB}}
-\frac{3}{16}\,{\frac {{B_{{r}}}
^{2}}{A{B}^{2}}}-\frac{1}{8}\,{\frac {B_{{r}}A_{{r}}}{{A}^{2}B}}
-{\frac {11}{16}}\,{\frac
{{A_{{t}}}^{2}}{{A}^{2}B}}+\frac{1}{16}\,{\frac {{A_{{r}}}^{2}}{{
A}^{3}}}-\sqrt {AB}
-{\frac{15}{16}}\,{\frac{{B_{{t}}}^{2}}{{B}^{3}}} +\frac{1}{2}\,{\frac
{B_{{tt}}}{{B}^{2}}} \\
G^{u}_{\ u} & = & \frac{1}{8}\,{\frac {A_{{t}}B_{
{t}}}{A{B}^{2}}}-{\frac {7}{16}}\,{\frac
{{B_{{t}}}^{2}}{{B}^{3}}}-\frac{1}{4} \,{\frac
{A_{{tt}}}{AB}}+\frac{1}{4}\,{\frac
{B_{{tt}}}{{B}^{2}}}+\frac{1}{16}\,{ \frac
{{A_{{t}}}^{2}}{{A}^{2}B}}-\frac{1}{8}\,{\frac
{B_{{r}}A_{{r}}}{{A}^{2}B }}-\frac{1}{16}\,{\frac
{{B_{{r}}}^{2}}{A{B}^{2}}}+{\frac {7}{16}}\,{\frac {{A
_{{r}}}^{2}}{{A}^{3}}}-\frac{1}{4}\,{\frac
{A_{{rr}}}{{A}^{2}}}+\frac{1}{4}\,{\frac { B_{{rr}}}{AB}}
\end{eqnarray}
and $G^{t}_{\ r}=-AB^{-1}G^{r}_{\ t}$, $G^{\phi}_{\ \phi}=G^{u}_{\
u}$. The subscripts $r$ and $t$ on the metric functions $A$ and
$B$ denote partial differentiation with respect to $r$ and $t$,
respectively.

We note that all terms in the expressions for $G^{r}_{\ t}$ and $G^{u}_{\ u}$ originate from the basic metric functions
$g_{rr}$ and $g_{tt}$ ($A$ and $B$).  The terms of the form $\sqrt{AB}$ in $G^{t}_{\ t}$ and $G^{r}_{\ r}$ arise from
derived `radial' metric functions (e.g., the term $\sqrt{AB}$ in equation (\ref{Gtt}) arises as a product of the metric
components $g_{tt}$, $g_{rr}$, $g_{uu}$ and $g_{\phi\phi}$; also see the comment in subsection III.A.3). In order to
avoid unnecessary complications in the averaging procedure we shall not deal with derived quantities directly
\footnote{This was pointed out to us by R. Zalaletdinov.}. Therefore, we eliminate the $\sqrt{AB}$ terms by considering

\begin{equation}
G^{t}_{\ t}-G^{r}_{\ r}=\frac{3}{4}\,{\frac {B_r A_r }{A^{2}B }}+{
\frac {7}{8}}\,{\frac {B_{t}^{2}}{B^{3}}+}{\frac {7}{8}}\,{\frac
{A_{r}^{2}}{A^{3}}}+\frac{3}{4}\,{\frac {A_t B_t}{A
B^{2}}}-\frac{1}{2}\,{\frac {A_{rr} }{A^{2}}}+{\frac
{7}{8}}\,{\frac {B_{r}^{2}}{A B^{2}}}+{\frac {7}{8}}\,{\frac
{A_{t}^{2}}{A^{2}B }}-\frac{1}{2} \,{\frac {B_{rr}}{A  B
}}-\frac{1}{2}\,{\frac {A_{tt} }{A B  }}-\frac{1}{2}\,{\frac
{B_{tt} }{B^{2}}}
\end{equation}
and use the contracted Bianchi identity to determine the remaining component of the Einstein tensor.  Taking averages,
we now obtain the appropriate form for the MG equations and hence the correlation tensor $C^{a}_{\ b}$.  For example,
we have that\footnote{$C^{a}_{\ b}$ is implicitly defined by this expression.  The formal definition of the correlation
tensor is given in \cite{Zala}.}

\begin{eqnarray}
C^{r}_{\ t} & \equiv & G^{r}_{\ t}\bra g \ket - \bra G^{r}_{\ t} \ket \nonumber\\
\label{ket}
 & = & -\frac{5}{8}\left[\frac{\bra B_r \ket \bra A_t \ket}{\bra A
\ket^2\bra B \ket}-\bra \frac{B_rA_t}{A^2B}\ket \right]
-\frac{7}{8}\left[\frac{\bra B_r \ket \bra B_t \ket}{\bra A \ket
\bra B \ket^2}-\bra \frac{B_rB_t}{AB^2}\ket \right]
-\frac{7}{8}\left[\frac{\bra A_r \ket \bra A_t \ket}{\bra A \ket^3}-\bra
\frac{A_rA_t}{A^3}\ket \right] \nonumber\\
& &-\frac{1}{8}\left[\frac{\bra A_r \ket \bra B_t \ket}{\bra A \ket^2\bra B \ket}-\bra \frac{A_rB_t}{A^2B}\ket \right]
+\frac{1}{2}\left[\frac{\bra A_{rt} \ket}{\bra A \ket^2}-\bra \frac{A_{rt}}{A^2}\ket \right]
+\frac{1}{2}\left[\frac{\bra B_{rt} \ket}{\bra A \ket\bra B \ket}-\bra \frac{B_{rt}}{AB}\ket \right].
\end{eqnarray}

\noindent In the above, by virtue of VPC, the average is now a simple average defined by

\begin{equation}
\label{bra}
\bra f(r,t) \ket \equiv
\frac{1}{TL}\int^{\frac{T}{2}}_{t^{\prime}=-\frac{T}{2}}
dt^{\prime} \int^{\frac{L}{2}}_{r^{\prime}=-\frac{L}{2}}
dr^{\prime}f(r+r^{\prime}, t+t^{\prime}),
\end{equation}

\noindent which, for smooth functions with a sufficiently slowly varying dependence on cosmological time,
essentially reduces to a spatial average in terms of the averaging scale  $L$ (see the next section).

It is instructive to consider the FLRW metric in VPC. The metric is given by (\ref{vpcss}), with
\begin{equation} \label{frwmetriccomps}
A  =  \frac{R^2}{F^4}; \quad B  =  \frac{1}{R^6}
\end{equation}
where $R= R(t)$ and $F=F(r)$, subject to $\frac{dF}{dr}=(\sqrt{1-kF^2})F^{-2}$, and $k=-1,0$ or $1$. The Einstein
tensor in these coordinates is given by
\begin{equation}
G^a_{\ b} = \left[
\begin{array}{cccc}
-3R^4\dot{R}^2 - 3kR^{-2} & 0 & 0 & 0\\
0 & -7R^4\dot{R}^2 - 2R^5\ddot{R}
- kR^{-2} & 0 & 0\\
0 & 0 & -7R^4\dot{R}^2 - 2R^5\ddot{R}
- kR^{-2} & 0 \\
0 & 0 & 0 & -7R^4\dot{R}^2 - 2R^5\ddot{R} - kR^{-2}
\end{array}
 \right].
 \end{equation}
The spatial curvature term is given by $G^a_{\ b} = {-k}{R_0^{-2}}
\mbox{{\it diag}}[3, 1, 1, 1]$  (whereby spatial curvature term we
mean the Einstein tensor corresponding to a spacetime with
constant spatial curvature), with effective equation of state
$p_{\mbox{eff}} = -\frac{1}{3} \rho_{\mbox{eff}}$.

\section{Inhomogeneous Spacetimes} The form of the correlation tensor
now depends on the assumed form for the inhomogeneous
gravitational field and matter distribution. Let us assume that

\begin{equation}
\label{Matter}
A(r,t)=\bra A(r,t) \ket \left[ 1+
\sum_{n=1}^{\infty}a_{n}(t)L^{n}\sin\left(\frac{2n\pi}{L}r\right)
+
\sum_{n=1}^{\infty}{\bar{a}}_{n}(t)L^{n}\cos\left(\frac{2n\pi}{L}r\right)
\right],
\end{equation}
\begin{equation}
\label{Inhom}
B(r,t)=\bra B(r,t) \ket \left[ 1+
\sum_{n=1}^{\infty}b_{n}(t)L^{n}\sin\left(\frac{2n\pi}{L}r\right)
+ \sum_{n=1}^{\infty}{\bar
{b}}_{n}(t)L^{n}\cos\left(\frac{2n\pi}{L}r\right) \right],
\end{equation}
where $r$ is a radial variable (and, strictly speaking, we are assuming that $r \leq L$).  The assumptions
(\ref{Matter}) and (\ref{Inhom}) constitute a spatial Fourier decomposition of the metric functions in which the
variation in the timelike direction is assumed small and the dominant source of inhomogeneity arises from a
spatial variation of the gravitational field (thus the 4-volume average effectively reduces, in this case, to a
smoothing on a spatial domain).  Note that the coordinates $t$ and $r$ appearing in (\ref{vpcss}) are not the
usual `time' and `radial' coordinates; however, the unit magnitude timelike coordinate basis vector has zero
vorticity, which implies the existence of a foliation of spacetime (where the $r$ coordinate parameterizes the
spatial hypersurfaces). Since the coordinate basis vectors $\partial_{t}$ and $\partial_{r}$ are independent
(i.e., the metric is diagonal), it follows that variation along timelike and spatial directions is not coupled.
Although other forms for the inhomogeneous gravitational field are possible (i.e., different assumptions to
(\ref{Matter},\ref{Inhom})), it is not expected that the main conclusions obtained in this paper will be
qualitatively affected. In equations (\ref{Matter}) and (\ref{Inhom}), $L\equiv L_{0}$ is treated as a parameter
and in the calculations that follow $L_{0}$ is effectively taken to be a small dimensionless parameter after a
renormalization of the variables using the speed of light (set to unity) and the present value of the Hubble
parameter, $H_{0}$, and a redefinition of the functions $a_{n}$, $b_{n}$.

With these assumptions\footnote{The calculations will depend on the assumed form for the inhomogeneities. The
assumptions (\ref{Matter}) and (\ref{Inhom}) are self-consistent and physically reasonable (at least at lowest
orders); see subsection III.B.2 for further discussion.} we have essentially assumed that averages effectively
become space averages with

\begin{equation}
\label{radial} \bra f(r) \ket =
\frac{1}{L}\int^{L/2}_{-L/2}f(r+r')dr'.
\end{equation}
We note that integrating the left-hand side of the first of these
equations (i.e., taking spatial averages) yields $\bra
A(r,t)\ket$. We also note that

\begin{eqnarray}
\label{partial} \frac{\partial}{\partial r}{A(r,t)} & = & \frac{\partial}{\partial r} {\bra A(r,t) \ket} \left
(1+\sum_{n=1}^{n=\infty}a_n L^n \sin\left(\frac{2n\pi}{L}r\right)
+\sum_{n=1}^{\infty}{\bar{a}}_{n}(t)L^{n}\cos\left(\frac{2n\pi}{L}r\right)\right)
\\
& + & \bra A \ket \bra \sum_{n=1}^{n=\infty}{2n\pi}a_n L^{n-1}
\cos\left(\frac{2n\pi}{L}r\right) -
\sum_{n=1}^{n=\infty}{2n\pi}a_n L^{n-1}
\sin\left(\frac{2n\pi}{L}r\right) \ket,
\end{eqnarray}
and
\begin{equation}
\bra {\frac{\partial}{\partial r}} {\bra A(r,t) \ket } \ket =\bra \frac{\partial}{\partial r}{A(r,t)} \ket,
 \quad \bra   \frac{\partial}{\partial t}\bra A(r,t) \ket   \ket=\bra \frac{\partial}{\partial t}{A(r,t)} \ket.
\end{equation}

\noindent Thus the assumed forms for the  inhomogeneous functions
$A$ and $B$ satisfy a set of appropriate and self-consistent
conditions.

We shall expand in powers of $L\equiv L_{0} <1$.  Since the coefficients $(a_1, a_2$, for example) are evolving
functions of time, these expansions may only be valid for a transient period of time. We calculate the correlation
tensor $C^{a}_{\ b}$ up to $\bo(L^2)$:

\begin{eqnarray}
C^{r}_{\ t} & = & 0 +  \frac{\pi}{8{{ \bra A \ket} }} \left[ \left\{ 3 \left( { b}_{{1}} -{ a}_{{1}}   \right) {{\dot
{\bar{b}}}}_{{1}}
 + 3 \left({ \bar{a}}_{{1}}  -{ \bar{b}}_{{1}}
\right){{\dot{b}}}_{{1}}  + \left(5\,{ b}_{{1}} -{ a}_{{1 }}\right) {{\dot{\bar{a}}}}_{{1}}  + \left( { \bar{a}}_{{1}}
-5\,{
\bar{b}}_{{1}} \right) {{\dot a}}_{{1}}\right\} \right. + \nonumber \\
& & \hspace{3cm} \left. \left\{ \left({ \bar{a}}_{{1}} {b}_{{1}}- {a}_{{1}}  { \bar{b}}_{{1}} \right) \left( 3
\frac{\bra B \ket_{t}}{{ \bra B \ket}} + 5 \frac { \bra A \ket{_t}}  {{ \bra A \ket}} \right) \right\} \right]L
+ \bo(L^2)  \\
C^{u}_{\ u} & = & \frac{\pi^2}{8{\bra A \ket}}{\left[\left(\bar{a}_{{1}}-\bar{b}_{{1}}\right)^2
+\left(a_{{1}}-b_{{1}}\right)^2-4\left(b^2_{{1}}+\bar{b}^2_{{1}}\right)\right]} + \frac{\pi}{8\bra A
\ket}\left(\bar{a}_{{1}}b_{{1}}-a_{{1}}\bar{b}_{{1}}\right) \left[ 3 \frac{\bra B \ket_{r}}{\bra B \ket} - \frac{\bra A
\ket_{r}}{\bra A \ket}\right]L+ \bo(L^2)  \\
C^{t}_{\ t}-C^{r}_{\ r} & = & \frac{\pi^2}{4{\bra A \ket}}{\left[\left(\bar{a}_{{1}}-\bar{b}_{{1}}\right)^2
+\left(a_{{1}}-b_{{1}}\right)^2-4\left(b^2_{{1}}+\bar{b}^2_{{1}}\right)\right]} + \frac{3\pi}{4\bra A
\ket}\left(\bar{a}_{{1}}b_{{1}}-a_{{1}}\bar{b}_{{1}}\right)\left[ \frac{\bra B \ket_{r}}{\bra B \ket} + \frac{\bra A
\ket_{r}} {\bra A \ket}\right]L + \bo(L^2)
\end{eqnarray}
The $\bo(L^2)$ terms have been calculated, but we have not explicitly displayed them here.

\subsection{Lowest Order Calculation}

From eqns. (7) and (\ref{partial}) we obtain, for example, $C^r\;
_t = 0 + \bo(L)$, and from eqns. (5) and (6)
\begin{equation}
\label{star}
C^a \; _b =
\left[\begin{array}{cccc}
C+ \frac{2 \ell}{\langle A \rangle} & 0 & 0 & 0\\
0 & C & 0 & 0\\
0 & 0 & \frac{\ell}{\langle A \rangle} & 0\\
0 & 0 & 0 & \frac{\ell}{\langle A \rangle}
\end{array}  \right] + \bo(L)
\end{equation}
where $C \equiv C^r \;_r$ and
$$ \ell(t) \equiv \frac{\pi^2}{8}\left[(a_1 -3b_1)(a_1 +b_1)+(\bar{a}_1 -3\bar{b}_1)(\bar{a}_1 +\bar{b}_1) \right].$$

We calculate $C$ from the contracted Bianchi identities.  We note
that if $C^a \; _b$ is isotropic (i.e., of the form of a perfect
fluid) then $C = \frac{\ell}{\langle A \rangle}$ and $C^a \; _b$
is of the form of a spatial curvature term.

\subsubsection{Bianchi Identities}

For the metric (\ref{vpcss}) and correlation tensor (\ref{star}),
to $\bo(L^0)$ the contracted Bianchi identities yield:
\begin{eqnarray}
&&
\label{dag}
C_r -\frac{1}{2} C\left( \frac{A_r}{A} + \frac{B_r}{B} \right)
+ \frac{1}{2} \frac{\ell}{\langle A \rangle} \left(\frac{A_r}{A} -
\frac{B_r}{B}  \right) =0  \\
\label{ddag} && C_t + \frac{2{\dot{\ell }}} {\langle A \rangle} -
\frac{1}{2} C\left(\frac{A_t}{A} + \frac{B_t}{B}  \right) -
\frac{1}{2} \frac{\ell}{\langle A \rangle} \left(\frac{A_t}{A} +
\frac{B_t}{B}  \right) = 0
\end{eqnarray}
where $C_r \equiv \frac{\partial C}{\partial r}$ and $C_t \equiv
\frac{\partial C}{\partial t}$.  The solution of these equations
(for C) depends on whether $B_r$ is zero or not.

\subsubsection{FLRW background}

In the case that $B_r =0$, as in the case of a FLRW background, equation (\ref{dag}) immediately yields $C \equiv
\frac{\ell}{\langle A \rangle}$ and $\langle A \rangle _{, r} =0$, and equation (\ref{ddag}) then yields
$\frac{\ell}{\langle A \rangle} = \ell_0 R^{-2}$ (where $\ell_{0}$ is a constant that depends on the spatial averaging
scale). Therefore, in this case we obtain
\begin{equation}
\label{stars}
 C^a \; _b = \ell_0 R^{-2}
\left[\begin{array}{cccc}
3 & 0 & 0 & 0 \\
0 & 1 & 0 & 0\\
0 & 0 & 1 & 0\\
0 & 0 & 0 & 1
\end{array} \right] \; ,
\end{equation}
 and $C^a \; _b$ is necessarily of the form of a spatial
 curvature term.

 In order to fully reconcile (\ref{star}) and (\ref{stars}), we note that
 since $A = R^2 F^{-4}$ in the FLRW case, we obtain
 \begin{equation}
 \label{FLRW}
 \frac{\ell}{\langle A \rangle} =
 \frac{\ell \langle F^4 \rangle}{R^2} = \frac{\ell_0}{R^2}
 = \frac{-k}{R^2}.
 \end{equation}
 Here $\langle A \rangle$ must be interpreted as an averaged
 spatial
 curvature, so that $\langle F^4(r) \rangle$ is replaced by
 $\langle 2F^2 F^2_r + F^3 F_{rr}\rangle$, which is constant. That is,
 the term $\langle A \rangle$ in the above equation must
 be interpreted correctly.

 \subsubsection{Non-FLRW background}

 If $B_r \neq 0$, then eqn. (\ref{dag}) can be integrated to obtain
 \begin{equation}
 C = -\frac{\ell}{\langle  A\rangle} + f(t)
 \frac{(AB)^{1/2}}{A^{2\ell}} \label{dddag}
  \end{equation}
(we note that, in general, this expression is different to what we would have obtained if we had averaged eqns.
(\ref{vpcss}), (\ref{Matter}) and (\ref{Inhom}),  with the  $\sqrt{AB}$ term, directly). Eqns (\ref{dddag}) and
(\ref{ddag}) then yield
 $$ \dot{\ell} = - \left[ \frac{2f}{A^{2 \ell}}  \right]_{, t}
 A^{3/2} B^{1/2}. $$
Since $A = A(r, t)$  $(B_r \neq 0)$, in general the solution of
this equation yields $\ell = \ell_0$ (constant) and $f(t) = g(r)
A^{2\ell}$, so that if $f  \neq 0$, $A(r,t)$ is separable.  In the
latter case, in general $B(r,t)$ is also separable, and the two
separate terms in (\ref{dddag}) can be of a comparable form.

We note that $C^a \; _b$ is necessarily anisotropic (and cannot be
formally equivalent to a perfect fluid). We can always write
\begin{equation}
C^a \; _b = \frac{\ell_0}{\langle A \rangle}
\left[\begin{array}{cccc}
3 & 0 & 0 & 0\\
0 & 1 & 0 & 0\\
0 & 0 & 1 & 0\\
0 & 0 & 0 & 1
\end{array}  \right]
-\Pi \left[\begin{array}{cccc}
1 & 0 & 0 & 0\\
0 & 1 & 0 & 0\\
0 & 0 & 0 & 0\\
0 & 0 & 0 & 0
\end{array}  \right]
\label{eqn.s}
\end{equation}
where $\Pi \equiv -\left\{g(r)\langle AB \rangle^{1/2} - \frac{2
\ell_0}{\langle A \rangle} \right\}$. The correlation tensor $C^a
\; _b$ automatically satisfies the contracted Bianchi identities
(\ref{dag}) and (\ref{ddag}). It can be interpreted as the sum of
a perfect fluid and an anisotropic fluid (when $B_r \ne 0$). For
an anisotropic fluid in spherically symmetric coordinates the
energy-momentum tensor is of the form $diag\left[ -\mu, p_{||},
p_\perp, p_\perp\right]$, where $p_{||} = p + \frac{2}{3}\pi$ and
$p_\perp = p -\frac{1}{3}\pi$, and $\pi$ is the anisotropic
pressure. From above, we see that if the (total) correlation
tensor $C^a \; _b$ is interpreted as an anisotropic fluid, it
follows that $\Pi = - \pi$ and $\mu = 3p$.

If both terms separately satisfy the contracted Bianchi identity, then the first term can be interpreted as a spatial
curvature term.  The second term can be interpreted as an anisotropic fluid with $p_\perp = 0$ and $p_{||} = -
\rho_{\mbox{eff}}$ (which is similar to the equation of state for a cosmological constant). For $f(t) = g(r) A^{2\ell}$
($g(r) \neq 0$) and $A \equiv \overline{F}^{-4} (r) \overline{R}^2 (t)$, we obtain $\rho_{\mbox{eff}} = 2 \ell_0
\overline{F}^4(r) R^{-2} = g(r) \overline{F}^{-2} \overline{R}(t) B^{1/2}$, so that $B$ is separable and of the form $B
= b(r) \overline{R}^{-6}(t)$ (compare with eqn. (\ref{frwmetriccomps})).  One solution gives $\rho_{\mbox{eff}} =0$, so
that in this case $C^a \; _b$ is of the form of a perfect fluid and is thus necessarily of the form of a spatial
curvature term.

Although the correlation tensor $C^a \; _b$ satisfies the
contracted Bianchi identities, when interpreted as the sum of a
spatial curvature perfect fluid and an anisotropic fluid through
(\ref{eqn.s}), the two separate fluid do not in general satisfy
separate conservation equations. However, eqn. (\ref{ddag}) can be
rewritten in the form of a conservation law for the anisotropic
pressure $\Pi$,

\begin{equation}
\label{cons} \Pi_t -\frac{1}{2} \Pi \left( \frac{A_t}{A} +
\frac{B_t}{B} \right) + \frac{\ell_0}{\langle A \rangle}
\left(2\frac{A_t}{A} + \frac{B_t}{B}  \right) =0,
\end{equation}
in VPC where the metric is given by eqn. (\ref{vpcss}) and $u^a$
is comoving to order $\bo(L^2)$ (note that the expansion and shear
are given by
\begin{equation}
\theta =-\frac{1}{2 B^{\frac{1}{2}}} \left( \frac{B_t}{B} \right), ~~ \sigma =\frac{1}{2
{\sqrt{6}}B^{\frac{1}{2}}}\left(3\frac{A_t}{A} + \frac{B_t}{B} \right),
\end{equation} respectively; compare this with eqn.
(\ref{condi:integrability}) in  Appendix C).

\subsubsection{Anisotropic Fluid}

The second term in the correlation tensor is of the form of an energy-momentum tensor for an anisotropic fluid,
with energy density $\mu$, a pressure $p_{||}$ parallel to the radial unit normal  and a perpendicular pressure
$p_\perp$\footnote{We emphasize that although the correlation tensor has the form of an anisotropic fluid (with
certain properties) it is not a material source but arises from the averaging procedure.  For example, the
conservation law (\ref{cons}) is simply a formal expression for the contracted Bianchi identities.}. Fluids with
an anisotropic pressure have been studied in the cosmological context for a number of reasons: an energy-momentum
tensor of this form arises formally if the source consists of two perfect fluids with distinct four-velocities, a
heat conducting viscous fluid under some circumstances, a perfect fluid and a magnetic field, and in the presence
of particle production \cite{Kras:1996,CT94,ZimPav}. In particular, the energy-momentum tensor of a cosmic string
\cite{Vilenkin} is of the form of an anisotropic fluid with $\mu=-p_{||}$, $p_\perp=0$ (such an equation of state
also arises in other early universe applications). Anisotropic fluids in spherically symmetric cosmological models
have been studied in \cite{CT94,aniso,Kras:1996}. In addition, the energy-momentum arising from the gravitational
field of a global monopole is formally an anisotropic fluid which is static and spherically symmetric
\cite{Vilenkin}. We also note that in an investigation of the consequences of an imperfect dark energy component
on the large scale structure, the effect of anisotropic perturbations (due to the dark energy) on the cosmic
microwave background radiation was studied. It was found that an anisotropic stress is not excluded by the
present day cosmological observations \cite{Koivisto}.

Let us comment on the astrophysical applications of an anisotropic
fluid. It is known that dark matter is a major constituent of the
halos of galaxies \cite{DM}. By an analysis of observed rotation
curves, under reasonable assumptions (e.g., that galaxies can be
modeled as spherically symmetric) it has been found that the dark
matter is of the form of an anisotropic fluid \cite{Lake}. This
has been taken up in \cite{Boon}, in which the consequences of
anisotropic dark matter stresses are discussed in the weak field
gravitational lensing (where it was noted that in any attempt to
model dark matter in galactic halos with classical fields will
lead to anisotropic stresses comparable in magnitude with the
energy density).

Finally, we note that for a 4-dimensional spacetime with a metric
of the form $g_{ab} = diag[-1,1,(1+cu^2)^{-1},(1+cu^2)]$ (in VPC),
where $c$ is a constant, we have that $G^a \; _b = diag[c,c,0,0]$.
The metric is of the product form $R^2 \times S^2$, and is
therefore the tensor product of a 2-dimensional flat space and
(for $c<0$) a 2-dimensional sphere (which are two 2-dimensional
spaces of constant curvature). Hence, the second term can also be
interpreted in terms of spatial curvature (although we again note
that each of the two terms, namely the spatial curvature term and
the anisotropic term, do not separately satisfy the contracted
Bianchi identity).

\subsection{Further calculations}

\subsubsection{Higher order terms}

We can consider the contribution of the higher order $\bo{(L)}$
and $\bo{(L^2})$ terms. In the case of a FLRW background (with
$B_r =0$, $\frac{\ell}{\langle A \rangle} = \ell_0 R^{-2}$, where
$\ell$ is constant), the $\bo{(L)}$ terms all vanish. We can see
from Appendix B that this occurs trivially when
$(\bar{a}_{{1}}b_{{1}}-a_{{1}}\bar{b}_{{1}})=0$, which occurs for
a sine series expansion (only) or a cosine series (only), or for a
single trigonometric series with a composite argument (using the
double angle formulae). The $\bo{(L^2)}$ contributions to $C^r_{\
t}$ then vanish, and the $\bo{(L^2)}$ contributions to the
remaining components of the correlation tensor are of the form of
a spatial curvature term (the $\bo{(L^2)}$ terms are not displayed
explicitly in the Appendix). That is, the effect of the higher
order terms is simply to renormalize the spatial curvature term.

In the non-FLRW background case, in general we must have
$(\bar{a}_{{1}}b_{{1}}-a_{{1}}\bar{b}_{{1}})=0$, and the higher
order terms do not play any significant role (as above).

\subsubsection{Discussion}

In equations (\ref{Matter}) and (\ref{Inhom}), $L$ is essentially treated as a dimensionless parameter, which is
sufficient to the lowest order of approximation.  In principle, in the cosmological setting $L$ depends on the
Hubble scale $H^{-1}$ and might also be related to a scale dependent on structure formation, both of which vary
with cosmological time. Therefore, in general, $L$ will be time dependent.

To lowest order we assume that $L=L_0$ is fixed and integration is taken over a comoving domain (and presently
$L_{0} \sim 10^{-1}$).  Assuming $L$ is time dependent, $L=L(t)$, we have that a typical correction term is of the
form
\begin{equation}
C_{T} \equiv \left\{1- \frac{1}{c_{n}(t)L^n} \frac{1}{T}\int_{-T/2}^{T/2} c_{n}(t)L^n dt \right\}.
\end{equation}
Assuming time evolution is of the order of the Hubble scale, we have that $C_T \sim {\mathcal O}(L_{0}\times H_{0}T)$,
where $H_{0}$ is the current value of the Hubble parameter.  Clearly such corrections are of order ${\mathcal O}(L_0)$
compared to the contributions calculated above.  Moreover, these corrections are negligible over small time averaging
scales $T$ (compared to $H_{0}$; i.e., $H_{0}T$ small).  In addition, we have that
\begin{equation}
\bra \frac{\partial}{\partial t}\bra f \ket \ket \cong \bra \frac{\partial}{\partial t}
\left\{\frac{1}{L}\int_{-L/2}^{L/2} f dr\right\} \ket \cong \bra \frac{\partial f}{\partial t} \ket + C_{\partial T},
\end{equation}
where
\begin{equation}
C_{\partial T} \equiv \bra \frac{1}{L}\frac{dL}{dt}\left\{ f-\bra f \ket \right\} \ket \sim L_{0}H_{0}\overline{C},
\end{equation}
where the term $\overline{C}$ in the particular case of inhomogeneities of the form (\ref{Matter})/(\ref{Inhom})
is negligible.  Therefore, $TC_{\partial T} \sim {\mathcal O}(L_0) \times H_{0}T$.  These corrections are
consequently of higher order and generally will only renormalize the spatial curvature term.

\subsubsection{Other inhomogeneous models}

The form of the correlation tensor depends on the assumed form for the inhomogeneous gravitational field and
matter distribution. We could consider alternatives to the form of the inhomogeneous metric
(\ref{Matter})/(\ref{Inhom}). We shall consider two alternative approaches here. First, we shall average an exact
inhomogeneous Lema\^{\i}tre-Tolman-Bondi solution. Second, we shall discuss a linear inhomogeneous perturbation of
an exact FLRW model.

However, the main conclusions of this section will not be affected; namely, in most applications of interest the
correlation tensor is of the form of a spatial curvature, but in general it is not even of the form of a perfect
fluid. Moreover, higher order corrections are not expected to lead to significant effects; e.g., they alone cannot
account for a current acceleration.

\section{Lema\^{\i}tre-Tolman-Bondi model}

Let us consider averaging an exact solution. The spherically
symmetric dust solution is the exact Lema\^{\i}tre-Tolman-Bondi
(LTB) model \cite{LTB,Kras:1996}, which can be regarded as an
exact inhomogeneous generalization of the FLRW solution. In the
dust LTB model, from the Gauss-Codazzi equations the Einstein
tensor has the form of a spatial curvature tensor on spacelike
hypersurfaces (which we recall is not the same as the projected
Einstein tensor). Various aspects of the averaging problem in LTB
spacetimes have been studied \cite{RasanenLTB,NamTan}.

The first step is to take the  LTB solution \cite{LTB,Kras:1996} and rewrite it in VPC.
This is done in Appendix B. We note that taking averages using VPC  is of interest in its own right, and is an
advantage in that averaging can now be done in both space and time (this will be discussed further in \cite{LTBAV}).
From the Appendix, we obtain
\begin{equation}
ds^2=-\left(1-\frac{U^2}{R^4}\right)dt^2-2\frac{U}{R^4}dtdx+\frac{dx^2}{R^4}+R^2\left[\frac{du^2}{1-u^2}
+(1-u^2)d\phi^2\right].    \label{ltbvpc}
\end{equation}
The constraints of the original \ltb metric become a defining
equation for $U(t,x)$ (\ref{conu}) and a differential equation for
$R(t,x)$  (\ref{conr}), which then ensures an exact dust solution
with density $G^{tt}(t,x)$. Since in VPC the velocity of the dust
flow is $u^{a}=\left(1,U(t,x),0,0\right)$, the Einstein tensor
components satisfy $G^{tx}=UG^{tt},  G^{xx}=U^{2}G^{tt},
G^{uu}=G^{\phi\phi}=0$ (see eqns. (\ref{einsvpc})).

In appendix B we explicitly construct the FLRW dust models in VPC.
The spatially flat ($E_{0}=0$) \frw model in \vpc\ is given by
(\ref{frwflat}). The spatially closed ($E_{0}<0$) \frw model in
\vpc\ is given in eqn. (\ref{frwapprox}) in terms of an expansion
(of $R$, $U$ and $G^{tt}$) about the spatially flat \frw model
with $E_{0}=0$  (a similar expression exists for the $E_{0}>0$
\frw model).

\subsection{A Perturbative Solution}
We shall assume that $t_{B}(r)$ is zero, which implies that the bang time is uniform and we are consequently
restricting our choice of LTB models to those with no decaying modes.  Such models are of interest at later times, and
particularly in the study of structure formation\cite{Silk}, and are suitable for our purposes here. We shall also
consider solutions of the \ltb metric in \vpc\ as perturbations about the spatially flat \frw model given in
(\ref{frwflat}). In this respect our approximate solution will be an expansion with respect to $E_{0}$ and we require
the Einstein tensor to have the form of (\ref{einsvpc}) (i.e., the form of dust, after truncation of terms of
${\mathcal O}(E_{0}^{2})$ or higher). We begin by making the following ansatz on the form of $R$
\begin{equation}
R(t,x)=R_{0}+\alpha_{1}x^a t^b E_0 + \alpha_2 x^c t^d E_{0}^2,
\label{Ransatz}
\end{equation}
\noindent where $\alpha_1$, $\alpha_2$, $a$, $b$, $c$ and $d$ are
constants to be determined from requiring the Einstein tensor has
the form of a dust solution up to order $E_0$.  Substituting
(\ref{conu}) into (\ref{conr}) gives a partial differential equation (PDE) involving only R, a
subsequent substitution by (\ref{Ransatz}) then shows that the
first non-trivial term in the PDE occurs at ${\mathcal
O}(E_{0}^{3})$.  At this stage we are only interested in a
perturbative solution of the PDE, therefore to obtain necessary
conditions for a dust solution we simply require that the
coefficient of $E_{0}^{3}$ vanish, which yields three cases
($a=1/3,-2/3$ or $1/3-b/2$).  Choosing $a$, we  can use eqns.
$(\ref{Ransatz})$ and $(\ref{conu})$ to obtain $U(t,x)$.
Calculating the Einstein tensor and requiring it have the form of
(\ref{einsvpc}) allows us to determine the remaining constants
(the details are discussed in \cite{LTBAV}).

If $a=1/3$, a number of subcases occur. A typical solution (e.g.,
$b=0$, $c=5/6$ and $d=-1$) gives rise to
$G^{tt}=4/(3t^2)-2\alpha_1 x^{-2/3}E_0 + {\mathcal O}(E_{0}^2)$
and $G^{\phi\phi}={\mathcal O}(E_{0}^2)$, so that the truncation
of $E_{0}^2$ and higher terms results in an Einstein tensor of the
form of dust (\ref{einsvpc}). Other solutions give rise to a
$G^{tt}$ with no $E_0$ terms (but containing higher orders of
$E_0$, with $G^{\phi\phi}$ beginning at these higher orders) and
 $G^{tt}$ and $G^{xx}$ components with no
${\mathcal O}(1)$ terms (and beginning with $E_0$ terms, whereas
the other components begin at higher orders of $E_0$). After
averaging the $(-2\alpha_1 x^{-2/3}E_0)$ contribution, we obtain a
correction term to the density which is independent of $t$. This
constant correction  in the dust model may be related to a
cosmological constant or to an anisotropic source (this will be
further investigated in \cite{LTBAV}).

If $a=-2/3$, then the resulting dust solutions are of the form of a flat or curved \frw model (the closed model is
given by (\ref{frwpos})). That is, the solution in this case is typically of the form of a flat \frw model with spatial
curvature corrections.

If $a=1/3-b/2$ (where $b\neq 0,2$, these cases are discussed
above), setting $c=-1/3-b$, $d=2(b+1)$ and $\alpha_2 = -
81\alpha_{1}^2 b(b-2)^2/(80(3b-1))$ gives a dust solution up to
order $E_0^2$. Unlike previous cases, here we have an arbitrary
power of $x$, and there are two free parameters $b$ and
$\alpha_1$; however, these solutions are not more general because
other free parameters are constrained.

We can obtain more general perturbative dust solutions containing more free parameters by a superposition of the
solutions discussed above. For example, one such solution  is found by adding a solution with $a=1/3$, $b=0$, $c=5/6$
and $d=-1$ and a solution with $b=-16/3$, $a=3$, $c=5$, $d=-26/3$ and $\alpha_2 = -1452\alpha_{1}^2/85$, to obtain a
solution of the form (to order $E_0^2$)
\begin{equation}
R(t,x)=\alpha_{0}R_0 + [\alpha_1 x^{1/3}+\beta_1
x^{3}t^{-16/3}]E_0 + [\alpha_2 x^{5/6}t^{-1}-
(1452/85)\beta_{1}^{2} x^{5}t^{-26/3}]E_{0}^2, \label{Rsuper}
\end{equation}
with corresponding forms for $U(t,x)$ and $G^{tt}$, thus giving
rise to another (more general) perturbative dust solution.
Therefore, we can construct perturbative LTB solutions which can
be interpreted as having both spatial curvature and constant
correction terms.

\section{Cosmological Perturbations}

\subsection{Backreaction}

The theoretical approach is to solve the full problem to obtain the equations satisfied by the averaged
quantities, without assuming a given background. An alternative but more practical approach is to assume a
spatially homogeneous and isotropic background and study the effect of perturbations (that is, the backreaction)
on this FLRW background \cite{Futa,Bild-Futa:1991,new2,Rasanen:2005}. The starting point is the Einstein equations
in an appropriately defined background\cite{new1,new2}. The Einstein and energy-momentum tensors are then expanded
in metric and matter perturbations up to second order. The linear equations are assumed to be satisfied, and the
spatially averaged remnants provide the new background metric which takes into account the backreaction effect of
linear fluctuations computed up to quadratic order. The backreaction has been studied for scalar gravitational
perturbations \cite{new2}, and it was found that the equation of state of the dominant infrared contribution to
the energy-momentum tensor which describes backreaction can take the form of a negative cosmological constant.
This has led to the speculation that gravitational backreaction may lead to a dynamical cancelation mechanism for
a bare cosmological constant \cite{MarBra}. Since, in the perturbative approach, the averaged Einstein tensor will
rapidly come to dominate over the correlation tensor, it has been argued that this might also explain the presence
of a source of late-time acceleration \cite{MarBra}.

The main aim is to investigate the effect of these perturbations on the local expansion rate and to see how it
might differ from the background expansion rate. In an important study, using a 3+1 split and the Zeldovich
approach and assuming inhomogeneous perturbations about a dust Einstein de Sitter (FLRW) background, the equations
governing the time dependence of the scale factor due to backreaction were obtained by spatial averaging
\cite{Bild-Futa:1991}. The metric perturbations were assumed small, even when the density contrast is large (much
larger than unity). It was found that the scale factor dependence on the correlation terms acts like a (negative)
spatial curvature term (and, curiously, that the age is greater than in the exact flat background FLRW model)
\cite{Bild-Futa:1991}. Typically perturbations are small corrections, but since they are time dependent they can
become larger, although likely vanishing asymptotically to the future \cite{Rasanen:2005}.

Therefore, the resulting correlation terms are simply of the form of a spatial curvature term in the linear
perturbation analysis. This is true in general, and is certainly true in spherically symmetric cosmological models.
There are, as mentioned earlier, problems with the perturbative approach. First, the perturbation scheme breaks down
when perturbations become significant and affect the background. Second, there are potentially gauge effects arising
from the  choice of hypersurface on which to do spatial averaging.

\subsection{Discussion}

Recent observations are usually interpreted as implying that the Universe is very nearly flat, currently
accelerating \cite{SN} and indicating the existence of dark matter and dark energy \cite{Weinberg}. A cosmological
constant or a negative pressure fluid (or quintessence field) are candidates for the dark energy. However, as
noted earlier, inhomogeneities can affect the dynamics and may significantly affect the expansion rate of the
spatially averaged ``background'' FLRW universe (the effect depending on the scale of the initial inhomogeneity)
\cite{Bild-Futa:1991}. Therefore, a more conservative approach to explain the acceleration of the Universe without
introduction of exotic fields is to utilize a backreaction effect due to inhomogeneities of the Universe.

It has been suggested that backreactions from inhomogeneities smaller than the Hubble scale could explain the
apparently observed accelerated expansion of the Universe today. This has been investigated by studying the
effective Friedmann equation describing an inhomogeneous Universe after smoothing out of the sub-horizon
cosmological perturbations, and it has been suggested that the acceleration in our Hubble volume might be possible
even if local fluid elements do not individually undergo accelerated expansion \cite{Kolb:2005,KMRnew}. However,
in \cite{Geshnizjani:2005} it was claimed that the perturbative effect proposed amounts to a simple
renormalisation of the spatial curvature, and in other work it has been argued that the acceleration cannot be
explained by the effects of inhomogeneities \cite{Flanagan:2005,IshWal}. However, after density fluctuations in
the Universe grow to be non-linear and begin to re-collapse, the perturbative expansion breaks down and reliable
results cannot be obtained beyond this based on perturbative calculations. More recently, a solution using the
gravitational backreaction of long wavelength (super-Hubble) fluctuation modes on the background metric was
presented \cite{MarBra}, and it was shown that in the presence of entropy fluctuations backreaction of the
non-gradient terms is physically measurable (compare with \cite{IshWal}).

In  further work \cite{Rasanen:2005} the relationship between backreaction and spatial curvature using exact
equations which do not rely on perturbation theory was studied in more detail, and it was argued that even though
the effect does not simply reduce to spatial curvature, the acceleration that results is accompanied by a growth
of spatial curvature to an extent that it is unlikely to be compatible with the CMB data. On the other hand, an
explicit example of an inhomogeneous Universe has been presented that leads to accelerated expansion after taking
spatial averaging \cite{NamTan}. The model universe is the LTB solution and contains both a region with positive
spatial curvature and a region with negative spatial curvature. It was found that after the region with positive
spatial curvature begins to re-collapse, the deceleration parameter of the spatially averaged universe becomes
negative and the averaged universe starts accelerated expansion. Further examples, in which the assumption of
spherical symmetry is relaxed, are discussed in \cite{buch2}.  In addition, inhomogeneities can lead to a
reinterpretation of the luminosity distance of cosmological sources in terms of its redshift, which may account
for the observed acceleration \cite{tomita}. However, it should be reiterated that there are subtleties when
dealing with spatially averaged quantities, even if the spatial averaging is over a limited domain, and that the
results discussed above may not apply to the quantities of physical interest \cite{buch2}. This point has also
been further emphasized in \cite{IshWal} (also see \cite{MarBra,KMRnew}).

\section{Conclusions}

We have calculated the form of the MG equations in the case of
spherical symmetry.  By working in VPC, we calculated the form of
the correlation tensor under some reasonable assumptions on the
form for the inhomogeneous gravitational field and matter
distribution. {\em The main result of this paper is that the
correlation tensor in a FLRW background must be of the form of a
spatial curvature, while for a non-FLRW background (with $B_r \ne
0$) the correlation tensor can be interpreted as the sum of a
spatial curvature and an anisotropic fluid.}  We note that working
in VPC (in which the determinant of the metric and hence the
volume element is constant) is useful in its own right,
particularly in the context of averaging.

The cosmological result that in the spherically symmetric case the averaged Einstein equations in an FLRW
background has the form of the Einstein equations of GR for a spatially homogeneous, isotropic macroscopic
spacetime geometry with an additional spatial curvature term, confirms the results in previous work in which we
were able to explicitly solve the MG equations to find a correction term in the form of a spatial curvature
\cite{CPZ}. The results of the calculations regarding averaging of an exact inhomogeneous LTB solution (presented
above), as well as calculations of linear perturbations (that is, the backreaction) in a spatially homogeneous and
isotropic background and the results of Buchert \cite{buch} also confirm and support this result.

The MG method adopted here is an exact approach in which inhomogeneities affect the dynamics on large scales
through correction terms\footnote{Hence the main criticisms of the backreaction approach to studying the possible
contributions to an accelerated expansion
\cite{Rasanen:2005,MarBra,Kolb:2005,Geshnizjani:2005,Flanagan:2005,IshWal} do not apply here.}. Averaging can have
a very significant dynamical effect on the evolution of the Universe; the correction terms change the
interpretation of observations so that they need to be accounted for carefully to determine if the models may be
consistent with an accelerating Universe. Averaging may or may not explain the observed acceleration. However, it
is clear that it cannot be neglected, and a proper analysis will not be possible without a comprehensive
understanding of the affects of averaging.

On cosmological scales (of the order of the inverse Hubble scale),
in which $B_r = 0$ and we have a FLRW background, averaging only
gives rise to a spatial curvature term. However, the effects of
averaging on astrophysical scales, such as galactic scales, are
also of interest. Under the assumption that a galaxy can be
approximated as spherically symmetric, where the background has
$B_r \ne 0$, averaging is found to give rise to a correlation
tensor of the form of an anisotropic fluid. This is of particular
interest since, as noted earlier, dark matter in galactic halos is
more accurately described by an anisotropic fluid \cite{Lake}.

In this paper we have also discussed  averaging in an inhomogeneous LTB solution. Writing the LTB solution in
volume preserving coordinates, we found a perturbative solution in which the correlation term can be interpreted
as the sum of a spatial curvature term and a constant correction term. We also discussed linear inhomogeneous
perturbations (that is, the backreaction) on an exact FLRW background. It was noted that the resulting correlation
terms are simply of the form of a spatial curvature term.

If the underlying microscopic spacetime has positive spatial curvature (as perhaps suggested by recent
observations \cite{Weinberg,SN,Spergel}), then we could obtain a cosmological model which is `closed' on local
scales, but as a result of the MG correlations behaves dynamically on macroscopically large scales as a flat
model, which might have considerable physical implications. Indeed, cosmological models which act like an Einstein
static model on the largest scales are possible even for models with zero or negative curvature on small scales;
thus at late times (and on the largest scale) a spatial curvature term will dominate the dynamics and the
correlations might stabilize the Einstein static model \cite{Ellis-Maartens} (also see Section 3.3 of
\cite{buch2}).

The MG analysis presented here is a self-consistent analysis. However, the MG approach is quite complicated and
can be difficult to implement in practice. Therefore, it is of interest to compare our results to the work of
Buchert \cite{buch2}. In this latter approach a 3+1 split of the equations is effected (which introduces some
gauge issues that presumably can be appropriately dealt with in the cosmological setting). More importantly, only
scalar quantities appear in the averaged equations. This implies some sort of `truncation' of the Einstein
equations in order for the equations to reduce to scalar equations. As a result, in general the Buchert equations
are not closed. In the approach taken here, the actual averages are constructed and therefore, in principle, the
forms for the averaged quantities take on a specific form. Consequently, our approach is more restrictive in the
sense that the system of governing equations is closed and so no further assumptions to close the system, which
may or may not be physical, are necessary. However, it is anticipated that it is possible to derive the averaged
scalar equations of Buchert \cite{buch} as some appropriate limit of MG. Whether any significant effects are
neglected in the Buchert approach could then be determined.

In Appendix C we show that in the spherically symmetric case the governing Einstein equations can reduce to equations
in terms of scalar quantities in some circumstances. Therefore, we may be able to compare our results with the work of
Buchert \cite{buch} in this case. Certainly, a spatial curvature term appears in Buchert's scheme when averaging in a
FLRW background, and it is expected that our analysis is consistent with the work of Buchert in a more fundamental
sense. Unfortunately, the perfect fluid case is very complicated for comparisons, and so it is more sensible to first
try to make the comparison in the dust case (i.e., the LTB model). Some brief comments are made in Appendix C and we
shall pursue this further in \cite{LTBAV}. Indeed, we note that for the models studied in section III, if the
correlation tensor is of the form of a perfect fluid then necessarily it is of the form of a spatial curvature (and
this result is thus trivially consistent with the work of Buchert). On the other hand, if the correlation tensor is of
the form of an anisotropic fluid, then a comparison with the work of Buchert is not possible.

\section{Appendix A: Macroscopic Gravity}

Let us review the spacetime averaging scheme adopted in macroscopic gravity (MG) \cite{Zala}. It is based on the
concept of Lie-dragging of averaging regions and is valid for any differentiable manifold. Choosing a compact
region $\Sigma \subset {\cal M}$ in an $n$-dimensional differentiable metric manifold $({\cal M}$, $g_{\alpha
\beta })$ with a volume $n$-form and a supporting point $x\in \Sigma $ to which the average value will be
prescribed, the average value of a geometric object, $p_\beta ^\alpha (x),\,x\in {\cal M\ }$, over a region
$\Sigma $ at the supporting point $x\in \Sigma $ is defined as
\begin{equation}
\label{defaver:MG}\overline{p}_\beta ^\alpha (x)=\frac 1{V_\Sigma
}\int_\Sigma {\bf p}_\beta ^\alpha (x,x^{\prime })\sqrt{-g^{\prime
}} d^nx^{\prime }\equiv \bra {\bf p}_\beta ^\alpha \ket \ ,
\end{equation}
where $V_\Sigma $ is the volume of the region $\Sigma $,
\begin{equation}
\label{volume}V_\Sigma =\int_\Sigma \sqrt{-g}d^nx\ ,
\end{equation}
the integration is carried out over all points $x^{\prime }\in
\Sigma $, $g^{\prime }=\det (g_{\alpha \beta }(x^{\prime }))$ and
the bold face object ${\bf p}_\beta ^\alpha (x,x^{\prime })$ is a
bilocal extension of the object $p_\beta ^\alpha (x)$,
\begin{equation}
\label{bilocext}{\bf p}_\beta ^\alpha (x,x^{\prime })={\cal
W}_{\mu ^{\prime }}^\alpha (x,x^{\prime })p_{\nu ^{\prime }}^{\mu
^{\prime }}(x^{\prime }) {\cal W}_\beta ^{\nu ^{\prime
}}(x^{\prime },x)\ ,
\end{equation}
by means of the bilocal averaging operator ${\cal W}_{\beta
^{\prime }}^\alpha (x,x^{\prime })$ and its inverse ${\cal
W}_\beta ^{\alpha ^{\prime }}(x^{\prime },x)$. The averaging
scheme is covariant and linear by construction, and the averaged
object $ \overline{p}_\beta ^\alpha $ has the same tensorial
character as $ p_\beta ^\alpha $. As a result of the coincidence
limit ($\lim _{x^{\prime }\rightarrow x}{\cal W}_{\beta ^{\prime
}}^\alpha (x,x^{\prime })=\delta _\beta ^\alpha$) and the
idempotency condition, the average tensor $\overline{p}_\beta
^\alpha (x)$ takes the same value as the original tensor $p_\beta
^\alpha (x) $ when the integrating region $\Sigma $  tends to
zero, which  implies that the averaging procedure commutes with
the operation of index contraction.

In order to obtain the averaged fields of geometric objects on
${\cal M}$ it is necessary to assign an averaging region $\Sigma
_x$ to each point $x$ of $ {\cal U}\subset {\cal M}$, where the
averaging integral is to be evaluated. To calculate directional,
partial and covariant derivatives of the averaged fields, regions
are related by Lie-dragging by means of a second bilocal operator,
which can also be taken to be ${\cal W}_\beta ^{\alpha ^{\prime
}}(x^{\prime },x)$ (which satisfies a divergence-free condition in
order for Lie dragging of a region to be a volume preserving
diffeomorphism) \cite{Zala}. The commutation relations simplify,
and the differential constraint for the idempotent bilocal reduces
to
\begin{equation}
\label{diffWW:1}{\cal W}_{[\beta ,\gamma ]}^{\alpha ^{\prime }}
+{\cal W}_{[\beta ,\delta ^{\prime }}^{\alpha ^{\prime }} {\cal
W}_{\gamma ]}^{\delta ^{\prime }}=0\,\ ,
\end{equation}
which has the general solution
\begin{equation}
\label{W}{\cal W}_\beta ^{\alpha ^{\prime }}(x^{\prime
},x)=f_i^{\alpha ^{\prime }}(x^{\prime })f{^{-1}}_\beta ^i(x)
\end{equation}
where $f_i^\alpha (x)\partial _\alpha =\mbox{\boldmath$f_{i}$}$ is
any vector basis satisfying the commutation relations $\left[
\mbox{\boldmath$f_{i}$},\mbox{\boldmath$f_{j}$}\right]
=C_{ij}^k\mbox{\boldmath$f_{k}$}$, with constant structure
functions $C_{ij}^k$. In any $n$-dimensional differentiable metric
manifold $({\cal M}$, $g_{\alpha \beta })$ with a volume $n$-form
there always exist locally volume-preserving divergence-free
operators ${\cal W}_\beta ^{\alpha^{\prime }}(x^{\prime },x)$ of
the form (\ref{W}) \cite{Zala}.

\subsection{Proper systems of coordinates}

 We can consider the MG averaging
scheme for a particular subclass of operators in which the averages and their properties are especially simple.
Such a coordinate system is the analogue for MG of the Cartesian coordinates in Minkowski spacetime \cite{MZ}. Let
us hereby restrict the class of solutions of the equations (\ref {diffWW:1}) to the subclass satisfying $\left[
\mbox{\boldmath$f_{i}$},\mbox{\boldmath$f_{j}$}\right] =0$; that is, $C_{ij}^k\equiv 0$. In this case the vector
fields $f_i^\alpha $ constitute a coordinate system and there always exist $n$ functionally independent scalar
functions $\phi ^i(x)$ such that the vector and corresponding dual 1-form bases are of the form
\begin{equation}
\label{bases:coor}f_i^\alpha (x(\phi^k)) =\frac{\partial x^\alpha
}{\partial \phi ^i} \,\ ,\quad f{^{-1}}_\alpha ^i(\phi(x^\mu))
=\frac{\partial \phi ^i}{\partial x^\alpha }\,\ .
\end{equation}
Thus, the bilocal operator ${\cal W}_\beta ^{\alpha ^{\prime
}}(x^{\prime },x)$ becomes
\begin{equation}
\label{W:coor}{\cal W}_\beta ^{\alpha ^{\prime }}(x^{\prime
},x)=\frac{
\partial x^{\alpha ^{\prime }}}{\partial \phi ^i}\frac{\partial \phi ^i}{
\partial x^\beta }\ .
\end{equation}
Since they are functionally independent, the set of $n$ functions
$\phi ^i(x)$ can be taken as a system of local coordinates on the
manifold ${\cal M}$, which will be called a proper coordinate
system \cite {MZ}. Therefore, in a proper coordinate system the
bilocal operator ${\cal W}_\beta ^{\alpha ^{\prime }}(x^{\prime
},x)$ takes the simplest possible form
\begin{equation}
\label{W:delta}{\cal W}_j^i(\phi ^{\prime },\phi )\equiv {\cal
W}_\beta ^{\alpha ^{\prime }}(x^{\prime },x)_{\mid x^\alpha =\phi
^i}= \mbox{\boldmath$\delta$}_\beta ^{\alpha ^{\prime }}\equiv
\mbox{\boldmath$\delta$}_j^i,
\end{equation}
where the bilocal Kronecker symbol $\mbox{\boldmath$\delta$}_\beta
^{\alpha ^{\prime }}$ is defined as
$\mbox{\boldmath$\delta$}_\beta ^{\alpha ^{\prime }}=\delta
_i^{\alpha ^{\prime }}\delta _\beta ^i$.

The definition of an average  consequently takes on a particularly simple form when written using a proper
coordinate system. The existence of volume-preserving bilocal operators ${\cal W}_\beta ^{\alpha ^{\prime }}$ of
this form  was proven in \cite{MZ}. Moreover, any proper coordinate system with a corresponding divergence-free
bivector is necessarily a system of volume-preserving coordinates.
In the case of a pseudo-Riemannian manifold the Christoffel symbols, $\Gamma _{\beta \alpha }^\alpha $, vanish and
partial differentiation and averaging commute in VPC.

Consequently, if the manifold $({\cal M}$, $g_{\alpha \beta })$ is a pseudo-Riemannian spacetime, the spacetime
averages defined in proper coordinates are Lorentz tensors, precisely like the averages in Minkowski spacetime.
The average value of a tensor field $p_\beta ^\alpha (x),\,x\in {\cal E}$, over a compact space region $S$ and a
finite time interval $\Delta t$ at a supporting point $(t,x^a)\in \Delta t\times S$ is thus
\begin{equation}
\label{defaver:ED} \bra p_\beta ^\alpha (t,x^a)\ket _{{\cal
E}}=\frac 1{\Delta tV_S}\int_{\Delta t}\int_Sp_\beta ^\alpha
(t+t^{\prime },x^a+x^{a\prime })dt^{\prime }d^3x^{\prime }\ ,
\end{equation}
where $V_S$ is the 3-volume of the region $S$, which is usually
taken as a 3-sphere of radius $R$ around the point $x^a$ at the
instant of time $t$, $ V_S=\int_Sd^3x^{\prime }$.

One issue of concern in the MG approach is the question of uniqueness; to what extent do spacetime averages depend
on the choice of the bilocal operator.  In the context of the present analysis, this raises the question of
whether the results obtained in this paper could depend on the choice of VPC (\ref{vpcss}).  It is clearly of
interest to study this question, and we hope to return to this in future work.  However, it is strongly
anticipated that the main conclusions of this paper will not be qualitatively affected by the choice of VPC;
namely, that in physical applications the correlation tensor is of the form of a spatial curvature (while in
general the correlation tensor is of the form of an anisotropic fluid).

\section{Appendix B: Lema\^{i}tre-Tolman-Bondi  Models}

A spherically symmetric  solution of the Einstein equation with
dust field  is given by the Lema\^{i}tre-Tolman-Bondi (LTB)
solution \cite{Kras:1996} with  metric
\begin{eqnarray}
  &ds^2=-d\tau^2+\frac{(R_{,r})^2}{1+2E(r)}dr^2+R^2d\Omega_2^2, \label{eq:metric}\\
  & \left(\frac{\dot
      R}{R}\right)^2=\frac{2E(r)}{R^2}+\frac{2M(r)}{R^3}, \label{eq:eq-R}
\end{eqnarray}
where $E(r)$ and $M(r)$ are arbitrary functions of $r$. The
solution of eqn. (\ref{eq:eq-R}) can be written parametrically by
using the variable $\eta=\int d\tau/R$,
\begin{eqnarray}
  &R (\eta,r)=\frac{M(r)}{-2E(r)}\left[1-\cos\left(\sqrt{-2E(r)}\,\eta\right)\right], \nonumber \\
  &t(\eta, r)=\frac{M(r)}{-2E(r)}\left[\eta-\frac{1}{\sqrt{-2E(r)}}\sin\left(\sqrt{-2E(r)}\,\eta\right)\right].
\end{eqnarray}
By introducing the following variables
\begin{equation}
  a(\tau,r)=\frac{R(\tau,r)}{r},\quad k(r)=-\frac{2E(r)}{r^2},\quad
  \rho_0(r)=\frac{6M(r)}{r^3},
\end{equation}
the metric and the evolution equation for the scale factor
$a(\tau,r)$ become
\begin{eqnarray}
  &ds^2=-d\tau^2+a^2\left[\left(1+\frac{a_{,r}r}{a}\right)^2
    \frac{dr^2}{1-k(r)r^2}+r^2d\Omega_2^2\right], \\
  &\left(\frac{\dot
      a}{a}\right)^2=-\frac{k(r)}{a^2}+\frac{\rho_0(r)}{3a^3}.
\label{eq:FReq}
\end{eqnarray}
Eqn.(\ref{eq:FReq}) is the same as the Friedmann equation with
dust, and we can regard the LTB solution as a model of an
inhomogeneous universe whose local behavior is equivalent to a
FLRW universe with a spatial curvature $k(r)$.

As a specific case,  Nambu and Tanimoto ~\cite{NamTan} assumed the
following spatial distribution of spatial curvature:
\begin{equation}
  k(r)=\frac{1}{L^2}\left[2\theta(r-r_0)-1\right],\quad 0\le r\le
  L,\quad 0\le r_0\le L
\end{equation}
and assumed that $\rho_0(r)=\rho_0=$constant. For $0\le r<r_0$,
the solution is that of a spatially open FLRW universe and for
$r_0<r\le L$, the solution is that of a spatially closed FLRW
universe.

\subsection{Volume Preserving Coordinate System}
Starting with the \ltb metric in the standard coordinate system $(\tau,r,\theta,\phi)$  above that is aligned with the
fluid flow, we obtain a volume preserving coordinate system (\vpc), $(t,x,u,\phi)$, by making the following coordinate
transformation
\begin{equation}
\begin{array}{ccc}
t=\tau, &
\displaystyle{x=\int\frac{R(\tau,r)^{2}R_{r}}{\sqrt{1+2E(r)}}dr},
& u=\cos\theta.  \label{vpctrans}
\end{array}
\end{equation}
Defining $U(t,x):=x_{\tau}$, and regarding $R=R(t,x)$, the line-element becomes
\begin{equation}
ds^2=-\left(1-\frac{U^2}{R^4}\right)dt^2-2\frac{U}{R^4}dtdx+\frac{dx^2}{R^4}+R^2\left[\frac{du^2}{1-u^2}
+(1-u^2)d\phi^2\right],     \label{ltbvpc}
\end{equation}
which has $g=-1$ as desired.  The constraints on the original \ltb metric ensuring a dust solution with density
\begin{equation}
G^{\tau\tau}=2\frac{M_{r}}{R(\tau,r)^2R_{r}},
\end{equation}
now become
\begin{equation}
U(t,x)=-\frac{2R_{t}R_{x}+RR_{tx}}{2R_{x}^{2}+RR_{xx}},
\label{conu}
\end{equation}
and
\begin{equation}
2(3R_{x}R_{t}-RR_{x}U_{x})U-2RR_{x}U_{t}+(7R_{x}^2+2RR_{xx})U^2-R_{t}^{2}-2RR_{tt}+R_{x}^{2}R^4=1.
\label{conr}
\end{equation}
Using eqn. (\ref{conu}), we can view equation (\ref{conr}) as a differential constraint for $R(t,x)$ (where $U(t,x)$ is
then derived once $R(t,x)$ is known).  As a result of eqns. (\ref{vpctrans}),(\ref{conu}) and (\ref{conr}), the
Einstein tensor has the following form
\begin{equation}
\begin{array}{llll}
G^{tt}=G^{\tau\tau}(t,x), & G^{tx}=UG^{tt}, & G^{xx}=U^{2}G^{tt}.
& G^{uu}=G^{\phi\phi}=0,   \label{einsvpc}
\end{array}
\end{equation}
These equations describe the corresponding \ltb dust solution in
volume preserving coordinates.  We note that the fluid was
comoving in the original coordinate system whereas in VPC the
velocity of the dust flow is $u^{a}=\left(1,U(t,x),0,0\right)$,
and hence is no longer comoving but is normalized,
$u_{a}u^{a}=-1$.

\subsection{Friedmann-Lema\^{i}tre-Robertson-Walker Cosmologies}

In order to compare with calculations we present the FLRW dust models in VPC.  In the original coordinate system the
\frw models have $E(r)=E_{0}r^2$, $M(r)=M_{0}r^3$ and a constant bang time $t_{B}$.  Setting
$L_{0}=|E_{0}|^{3/2}/M_{0}$ throughout, the spatially flat ($E_{0}=0$) \frw model in \vpc\ is given by

\begin{eqnarray}
R(t,x)=(3x)^{1/3}, &\  & U(t,x)=\frac{2x}{t-t_{B}},
\label{frwflat}
\end{eqnarray}
giving an Einstein tensor component (the other components are
determined from (\ref{einsvpc}))
\begin{equation}
G^{tt}=\frac{4}{3(t-t_{B})^2}.
\end{equation}

The exact spatially closed ($E_{0}<0$) \frw model in \vpc\ is
given by
\begin{equation}\label{frwpos}
\begin{array}{cc}
\displaystyle{R(t,x)  =
\frac{1}{2\sqrt{2}L_{0}}\sin\theta\left(1-\cos\eta\right)}, &
\displaystyle{U(t,x)=\frac{6\sqrt{2}L_{0}x\sin\eta}{(1-\cos\eta)^2}}, \\
\eta-\sin\eta  =  2\sqrt{2}L_{0}(t-t_B), &
\displaystyle{\theta-\sin\theta\cos\theta  =
\frac{32\sqrt{2}L_{0}^{3}x}{(1-\cos\eta)^3}},
\end{array}
\end{equation}
where $\eta=\eta(t)$ and $\theta=\theta(t,x)$.  The resulting
Einstein tensor component is
\begin{equation}
G^{tt}=\frac{48L_{0}^2}{(1-\cos\eta)^3}.
\end{equation}

\noindent  It is of interest to consider the form of $R$, $U$ and
$G^{tt}$ in the closed \frw model written as a perturbation about
the spatially flat \frw model. Expanding about $E_{0}=0$ and
defining $R_{0}=(3x)^{1/3}$, we obtain

\begin{eqnarray}\label{frwapprox}
R(t,x) & = &
R_{0}\left(1+\frac{12^{1/3}x^{2/3}}{15M_{0}^{2/3}(t-t_{B})^{4/3}}E_{0}-\frac{2^{1/3}x^{2/3}(19\cdot3^{2/3}x^{2/3}+
126(t-t_{B})^2)}{1575M_{0}^{4/3}(t-t_{B})^{8/3}}E_{0}^{2}\right)+{\mathcal O}(E_{0}^{3}), \\
U(t,x) & = &
\frac{2x}{t-t_{B}}+\frac{6^{2/3}x}{5M_{0}^{2/3}(t-t_{B})^{1/3}}E_{0}-\frac{39\cdot
6^{1/3}x(t-t_{B})^{1/3}}{175M_{0}^{4/3}}E_{0}^{2}+{\mathcal O}(E_{0}^{3}), \\
G^{tt} & = &
\frac{4}{3(t-t_{B})^2}-\frac{2\cdot6^{2/3}}{5M_{0}^{2/3}(t-t_{B})^{4/3}}E_{0}
+\frac{102\cdot6^{1/3}}{175M_{0}^{4/3}(t-t_{B})^{2/3}}E_{0}^2+{\mathcal
O}(E_{0}^{3}).
\end{eqnarray}
In the above perturbation scheme we have assumed that $E_0$ (and
$x$ and $t$) are small. The spatially open ($E_{0}>0$) \frw model
in \vpc\ can also be displayed (it is similar to the closed case
but with hyperbolic trigonometric functions replacing
trigonometric functions and appropriate sign changes-- see
\cite{LTBAV}).

\section{Appendix C: Spherically symmetric models in the 1+3 Formalism}

Using the Uggla and van Elst 1+3 formalism \cite{EU} for a perfect
fluid energy-momentum tensor, in the case of spherical symmetry we
have the evolution equations
\begin{eqnarray}
\dot{\theta} &=& - \,\frac{1}{3}\,\theta^{2} + ({\bf e}_{1} +
\dot{u} - 2\,a)\, (\dot{u}) - \frac{2}{3}\,(\sigma_{+})^{2}
- \frac{1}{2}\left(\mu+3p\right) + \Lambda \\
\label{tfrac13} \dot{\sigma_{+}} &=& - \,\theta\,\sigma_{+}
- ({\bf e}_{1}+\dot{u}+a)\,(\dot{u}) - {}^*\!S_+  \\
\dot{a} &=&  - \,\frac{1}{3}\,(\theta+\sigma_{+})\, (a+\dot{u})
\\
\label{tfrac23} \dot{{}^2\!K} &=&  -
\,\frac{2}{3}\,(\theta+\sigma_{+})\,{}^2\!K \\
\dot{\mu} &=&  - \,\left(\mu+p\right)\theta
\end{eqnarray}
and the Friedmann constraint
\begin{eqnarray}
0 &=& \frac{1}{3} \theta^2 + \frac{1}{2} {}^*\!R - \frac{1}{3}\sigma_+^2 -
\mu - \Lambda
\end{eqnarray}
(and the spatial constraint equations), where ${}^2\!K(t,x)$ is
the 2-curvature of the spheres. $\dot{u}$ is specified by choosing
a temporal gauge, and $p$ is specified by the fluid model.

There is only a single shear component, $\sigma \equiv \sigma_+$,
so that in the spherically symmetric case the shear can be
described by a single scalar.  The curvature, in some cases, can
also be described by a scalar.  Therefore, in the spherically
symmetric case we may be able to use scalar equations in some
appropriate limit to describe the model. In particular, choosing a
gauge in which $\dot{u} = 0$ and defining
\begin{eqnarray*}
R_d & = & ^2K\\
Q_d & = & - \frac{2}{3} \sigma^2
\end{eqnarray*}
we obtain from the Friedmann constraint
$$ 0 = \frac{1}{3} \theta^2 + \frac{1}{2} R_d + \frac{1}{2}
Q_d - \mu - \Lambda + 2 e_1(a) - 6a^2. $$ The evolution eqns.
(\ref{tfrac13}) and (\ref{tfrac23}) yield
\begin{eqnarray*}
2 \theta Q_d + \dot{\theta}_d & = & \sqrt{\frac{2}{3}}
Q^{1/2}_d R_d\\
\dot{R}_d + \frac{2}{3} \theta R_d & = & - \sqrt{\frac{2}{3}}
Q^{1/2}_d R_d
\end{eqnarray*}
which yields
$$ a^{-2}_d [ R_d a^2]^{\bullet} + a^{-6}_d [Q_d a^6]^\bullet = 0 $$
where $\theta \equiv 3\dot{a}_d/a_d$. These equations are valid
for both averaged and non-averaged scalar quantities.

We can see that in the case of spherical symmetry, the resulting
governing equations are equations for scalar quantities under some
circumstances. Therefore, we may be able to compare our results
with the work of Buchert \cite{buch} in this case. Certainly
Buchert \cite{buch} can obtain a spatial curvature term in an FLRW
background, and it is expected that our analysis is consistent
with the work of Buchert in a deeper sense. Unfortunately, the
perfect fluid case is very complicated for comparisons, and so it
is more sensible to try to make the comparison in the dust case
(i.e., the LTB model).

\subsection{The Dust Case}

Let us consider an inhomogeneous universe with irrotational dust.
Following Buchert \cite{buch}, one then obtains from Einstein's
equations the following equations of motion for the effective
scale factor, $a_\D$,

\begin{eqnarray}
  3\frac{\ddot a_\D}{a_\D}
  &=& -\frac{\kappa^2}{2} \bra \rho \ket_\D + Q_\D \,,
\label{eq:evolve}
\\
  3 \left( \frac{\dot a_\D}{a_\D} \right)^2
  &=& \kappa^2\bra \rho \ket_\D  -\frac{1}{2} {\cal R}_\D -
\frac{1}{2} Q_\D \,, \label{eq:friedman}
\end{eqnarray}
together with
\begin{equation}
 \Tdot{\left( a_\D^6 Q_\D \right)}
 + a_\D^4
 \Tdot{\left(a_\D^2  {\cal R} _\D \right)} =0 \,.
\label{condi:integrability}
\end{equation}
Here ${\cal R}$ is the spatial scalar curvature (not necessarily isotropic)  and
\begin{equation}
  Q_\D \equiv \frac{2}{3} \left( \langle \theta^2 \rangle_\D
               - \langle \theta \rangle_\D^2 \right)
               - \langle \sigma_{ij}\sigma^{ij} \rangle_\D \,.
\end{equation}

From the analysis of the spherically symmetric models we see that, in principle, we can obtain precise evolution
equations for ${\cal R}_\D, Q_\D $ (and, more generally, for the additional terms in the averaged equations for perfect
fluid models). Indeed, since we actually take averages in our analysis we will obtain specific forms for the averaged
quantities (i.e., in principle we will obtain explicit expressions for ${\cal R}, Q_\D $). We intend to study this in
the LTB models in more detail elsewhere \cite{LTBAV}.

However, in an analysis of the spherically symmetric collapse of dust in the Newtonian regime (assuming an irrotational
velocity field of the form ${\mathbf v}=v(r){\mathbf e}_{r}$ and other physical restrictions) \cite{Newt}, it was found
that the backreaction, $Q_{D}$, vanishes (as might be expected in the Newtonian approximation).  Therefore, in this
case the only effect of averaging in the Buchert approach is through a spatial curvature term. Unfortunately, the
Newtonian limit is not expected to capture the relativistic backreaction; a GR treatment is needed to discuss the
global effects of averaging.

In our relativistic MG approach let us assume a Newtonian-like 4-velocity field of the form
\begin{equation}
u^{a} = \frac{1}{\sqrt{B-Av^2}}[1,v,0,0],
\end{equation}
where $A$ and $B$ are the metric functions and $v=v(r)$ is assumed to be small (i.e., $v \ll 1$).  The corresponding
4-acceleration is then given by
\begin{equation}
A^{a}=-\frac{F}{2(B-Av^2)^2}\left[\frac{v}{B},\frac{1}{A},0,0\right],
\end{equation}
where
\begin{equation}
F \equiv AA_{t}v^3 + (2AB_r - A_{r}B)v^2 + (AB_{t}-2A_{t}B-2ABv_{r})v-BB_{r}.
\end{equation}
If we assume that the fluid is pressure-free (i.e., dust), then the acceleration is zero.  To lowest order in $v$, this
implies that $B_{r}=0$, and from earlier we conclude that the correlation tensor is of the form of a spatial curvature.
Consequently, in this approximation the Buchert approach \cite{Newt} and the MG approach are consistent.

{\em Acknowledgements}.

We would like to acknowledge discussions with Roustam Zalaletdinov
at the beginning stages of this work. We would like to thank
Thomas Buchert for many helpful discussions, and Robert van den
Hoogen for comments on the manuscript. This work was supported, in
part, by NSERC.

\end{document}